\documentclass[amssymb,epsf,11pt]{article}
\usepackage[colorlinks,urlcolor=black,linkcolor=blue,anchorcolor=blue,citecolor=blue]{hyperref}
\usepackage{epsf}
\usepackage{color}
\usepackage{amsmath}
\usepackage{graphicx}% Include figure files
\usepackage{dcolumn}% Align table columns on decimal point
\usepackage{bm}% bold math

\newcommand{\be}{\begin{equation}}
\newcommand{\ee}{\end{equation}}

\begin{document}
\title{\bf \Large Multifractal temporally weighted detrended partial cross-correlation analysis to quantify intrinsic power-law cross-correlation of two non-stationary time series affected by common external factors}

\author{\centerline{Bao-Gen Li$^1$,  Dian-Yi Ling$^1$~and~ Zu-Guo Yu$^{1,2}$\thanks{ Corresponding author, yuzuguo@aliyun.com}}\\
{\small $^{1}$Key Laboratory of Intelligent Computing and Information Processing of Ministry of Education }\\ {\small and Hunan Key Laboratory for Computation and Simulation in Science and Engineering,}\\
{\small Xiangtan University, Xiangtan, Hunan 411105, China.}\\
{\small $^{2}$School of Electrical Engineering and Computer Science, Queensland University of Technology,}\\
 {\small GPO Box 2434, Brisbane, Q4001, Australia.}}

\date{ }% It is always \today, today,
\maketitle           %  but any date may be explicitly specified

\begin{abstract}
When common factors strongly influence two cross-correlated time series recorded in complex natural and social systems, the results will be biased if we use multifractal detrended cross-correlation analysis (MF-DXA) without considering these common factors. Based on multifractal temporally weighted detrended cross-correlation analysis (MF-TWXDFA) proposed by our group and multifractal partial cross-correlation analysis (MF-DPXA) proposed by Qian {\it et al.}, we propose a new method---multifractal temporally weighted detrended partial cross-correlation analysis (MF-TWDPCCA) to quantify intrinsic power-law cross-correlation of two non-stationary time series affected by common external factors in this paper. We use MF-TWDPCCA to characterize the intrinsic  cross-correlations  between the two simultaneously recorded time series by removing the effects of other  potential time series. To test the performance of MF-TWDPCCA, we apply it, MF-TWXDFA and MF-DPXA on simulated series. Numerical tests on artificially simulated series demonstrate that MF-TWDPCCA can accurately detect the intrinsic cross-correlations for two simultaneously recorded series. To further show the utility of MF-TWDPCCA, we apply it on time series from stock markets and find that there exists significantly multifractal power-law cross-correlation between stock returns. A new partial cross-correlation coefficient is defined to quantify the level of intrinsic cross-correlation between two time series.
\end{abstract}
% PACS, the Physics and Astronomy Classification Scheme.
{\bf keywords:}Partial cross-correlation; multifractal temporally weighted detrended partial cross-correlation analysis (MF-TWDPCCA); partial cross-correlation coefficient.%Use showkeys class option if keyword display desired
\maketitle

%\begin{quotation}
%In recent years, as an effective tool, multifractal cross-correlation analysis has been widely used to study non-stationary time series in various fields. There are more and more examples
%that two time series are affected by the common forces, so we need to pay attention to multifractal partial cross-correlation analysis between time series. In this paper, a new method---multifractal temporally weighted detrended partial cross-correlation analysis (MF-TWDPCCA) is proposed, which is based on multifractal temporally weighted detrended cross-correlation analysis (MF-TWXDFA) and multifractal partial cross-correlation analysis (MF-DPXA). In this method, considering the sign problem of the fluctuation function of the detrended partial cross-covariance function, we estimate the local trend of the non-stationary time series with the geographically weighted regression from geoscience. In order to test the performance of the MF-TWDPCCA, we use it, MF-TWXDFA and MF-DPXA on three sets of simulated series. Numerical simulation demonstrate that MF-TWDPCCA has better performance than MF-TWXDFA and MF-DPXA. To further show its utility, MF-TWDPCCA is used to study the time series from the stock market, and we find that there obviously exists multifractal power law cross-correlation between stock returns.
%\end{quotation}

\section{Introduction}

\ \ Complex systems with interacting constituents exist in all aspects of nature and society, such as geophysics~\cite{Campillo2003}, solid state physics, climate system, ecosystem, financial system~\cite{Auyang1998, Plerou2008}, etc. In order to study the micro-mechanisms of these complex systems and their operating mechanisms in a statistical sense, people record and analyze the time series of observable quantities. The study of these time series will help us to understand things correctly, grasp the laws of nature, and make scientific decisions. There are relatively mature models and methods for the study of stationary time series, but unfortunately, the time series of real-world complex systems are usually non-stationary and possess long-range power-law cross correlations. Examples of this nature that have been reported include econophysical variables~\cite{LinDC2008,PodobnikB2009,SiqueiraJrEL2010,WangY2010,HeLY2010}, traffic signals~\cite{ZhaoX2011} and traffic flows~\cite{XuN2010}, brain activity and heart rate variability in healthy humans~\cite{LinDC2010}, topographic indices and crop yield in agronomy~\cite{KravchenkoAN2000,ZelekeTB2004}, self-affine time series of taxi accidents~\cite{ZebendeGF2011}, wind patterns and land surface air temperatures~\cite{JimenezHorneroFJ2011}, nitrogen dioxide and ground-level ozone~\cite{JimenezHorneroFJ2010}, the temperature and concentration fields of turbulent flows embedded in the same space as joint multifractal measures~\cite{AntoniaRA1975,MeneveauC1990}, sunspot numbers and river flow fluctuations~\cite{HajianS2010} and temporal and spatial seismic data~\cite{ShadkhooS2009}.

So far, various methods have been proposed for the long-range power law relationship of the cross-correlation between two non-stationary time series. In order to study the correlation between two time series, in 2008, Podobnik and Stanley proposed a Detrended Cross-Correlation Analysis (DCCA)~\cite{PodobnikB2008} algorithm based on the Detrended Fluctuation Analysis (DFA)~\cite{PengCK1994}. This provides a basis for succeeding study of cross-correlation. Since then, this method was continuously promoted and improved. In the same year, Zhou extended DCCA to multifractal case and proposed a Multifractal Detrended Cross-Correlation Analysis (MF-DXA)~\cite{ZhouWX2008} to study the multifractal cross-correlation  property of two non-stationary time series. O\'{s}wi\c{e}cimka {\it et al.} found that in the MF-DXA algorithm, taking the absolute value on the fluctuation function may lead to pseudo-correlation. That is, time series that do not have cross-correlation by themselves, but previous methods gave wrong results of existing cross-correlations between some time series. To solve this problem, O\'{s}wi\c{e}cimka {\it et al.} proposed a multifractal detrended cross-correlation analysis (MFDCCA)~\cite{Pswiecimka2014}. This method introduces the sign of fluctuation function when calculating generalized moments. In 2010, Zhou and Leung proposed the multifractal sliding window detrending fluctuation analysis (MF-MWDFA) and the multifractal temporally weighted detrended fluctuation analysis (MF-TWDFA)~\cite{ZhouYLeungY2010} for single time series. The innovation of MF-TWDFA is that it uses the idea of temporal weighting  to remove the local trend in the sliding window, which can avoid the sharp oscillation of the crossing position. Based on MF-TWDFA and MFDCCA, for analyzing the cross-correlation of multifractals of two time series, our group (Wei {\it et al.}) introduced the sign of the fluctuation function and adopted a Geographically Weighted Regression Model (GWR)~\cite{LeungYetal2000} to  propose a new method--- multifractal temporally weighted detrended cross-correlation analysis (MF-TWXDFA) ~\cite{WeiYL2017} in 2017.

If the two non-stationary time series to be studied are driven by a common third-party force or by common external factors, then the power law relationship of the cross-correlation between them observed by the method mentioned above may not reflect by their intrinsic relationship~\cite{KenettDY2009,KenettDY2010,ShapiraY2009}. Fortunately, Baba {\it et al.}~\cite{BabaK2004} found that if two time series affected by the external factors are additive, the levels of intrinsic cross-correlation between two time series can be measured by the partial cross-correlation coefficient. Yuan {\it et al.}~\cite{YuanN2015} focused on the detrended partial cross-correlation analysis (DPXA) coefficient in complex systems and gave a general calculation method. Almost at the same time, Qian {\it et al.}~\cite{QianXY2015} gave a general framework for DPXA and multifractal detrended partial cross-correlation analysis (MF-DPXA) for additive models.

In this paper, based on MF-TWXDFA and MF-DPXA, we propose a new method---multifractal temporally weighted detrended partial cross-correlation analysis (MF-TWDPCCA) to quantify intrinsic power-law  cross-correlation of two non-stationary time series affected by common external factors. Compared with MF-TWXDFA, our new method MF-TWDPCCA removes the influence of common factors and then can accurately detect the intrinsic cross-correlations of the two time series in the additive model. Compared with MF-DPXA, we use the idea of MF-TWXDFA to avoid the possibility of the pseudo-correlation.

\section{Multifractal temporally weighted detrended partial cross-correlation
analysis}

%%%%%%%%%%%%%%%%%%%%%fig1
\begin{figure*}[!htb]

\centering
  \begin{minipage}{6.5cm}
  \centering
  \includegraphics[width=6.5cm,height=6.5cm]{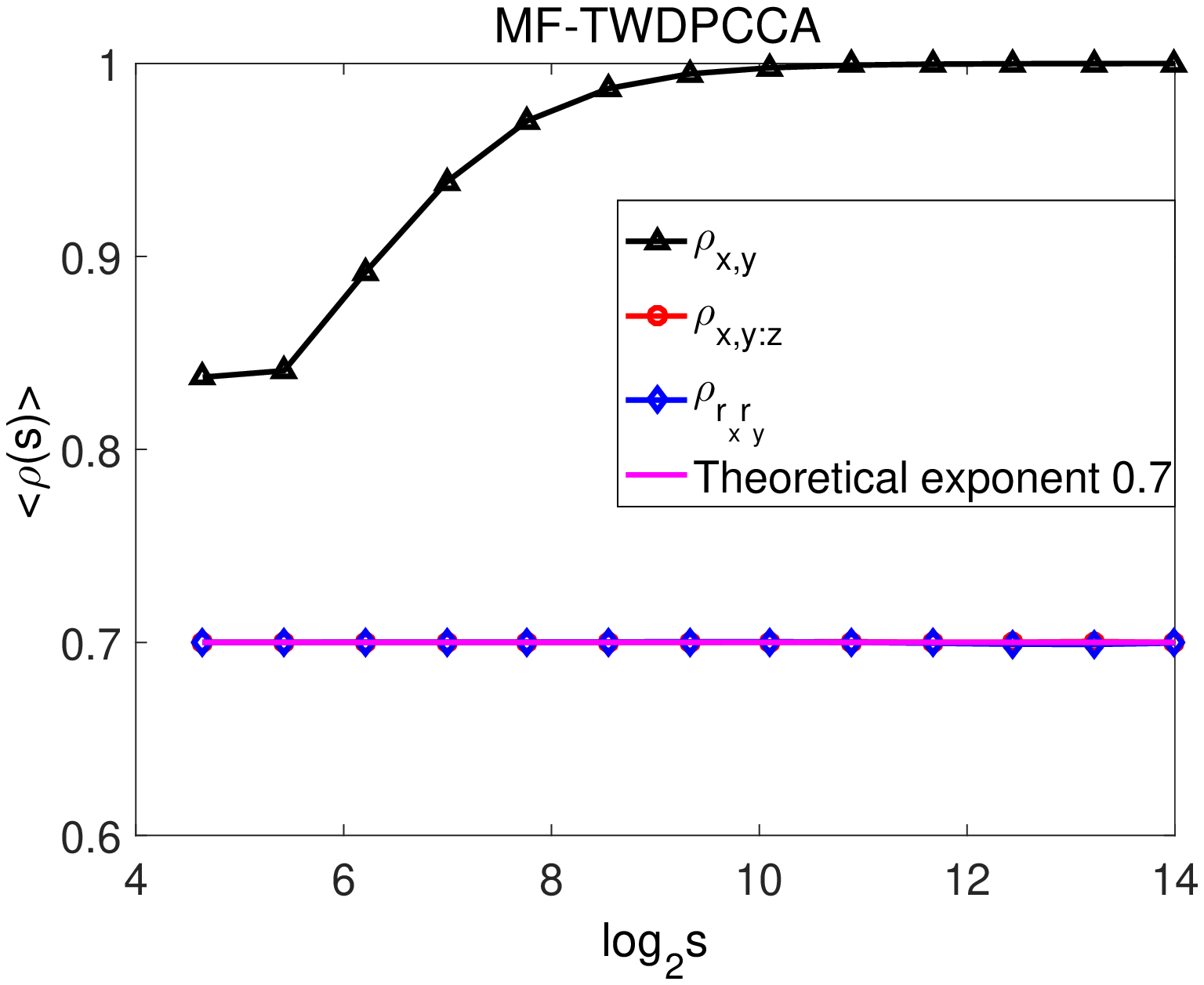}
  \end{minipage}
  \begin{minipage}{6.5cm}
  \centering
  \includegraphics[width=6.5cm,height=6.5cm]{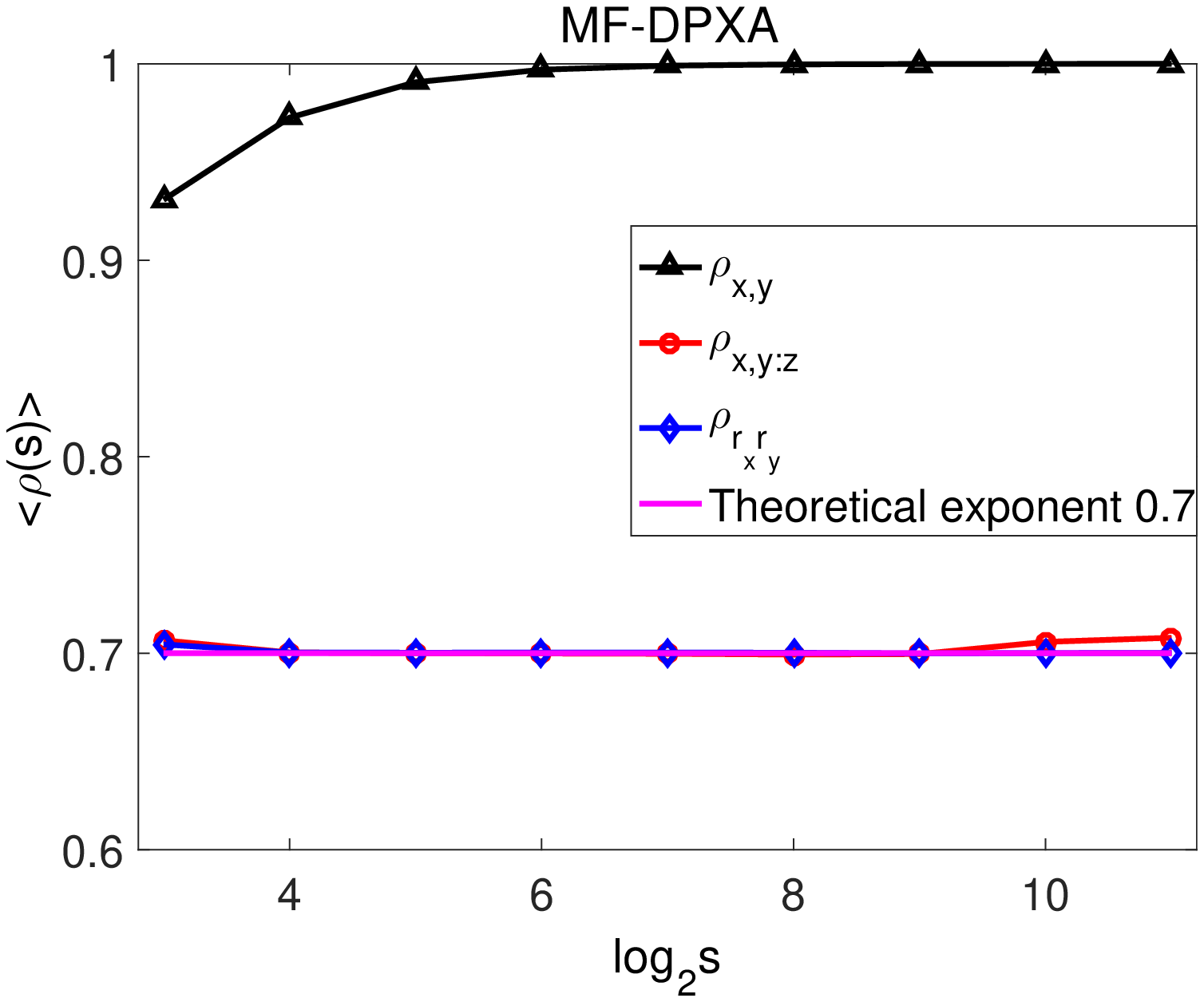}
  \end{minipage}
\caption{(Left) The detrended partial cross-correlation coefficients calculated by MF-TWDPCCA. (Right) The detrended partial cross-correlation coefficients calculated by MF-DPXA. Here $<\cdot>$ means the average over 100 realizations.}\label{rhocompaerd}
\end{figure*}

%%%%%%%%fig2
\begin{figure*}[!htb]
\centering
  \begin{minipage}{7cm}
  \centering
 \includegraphics[width=7cm]{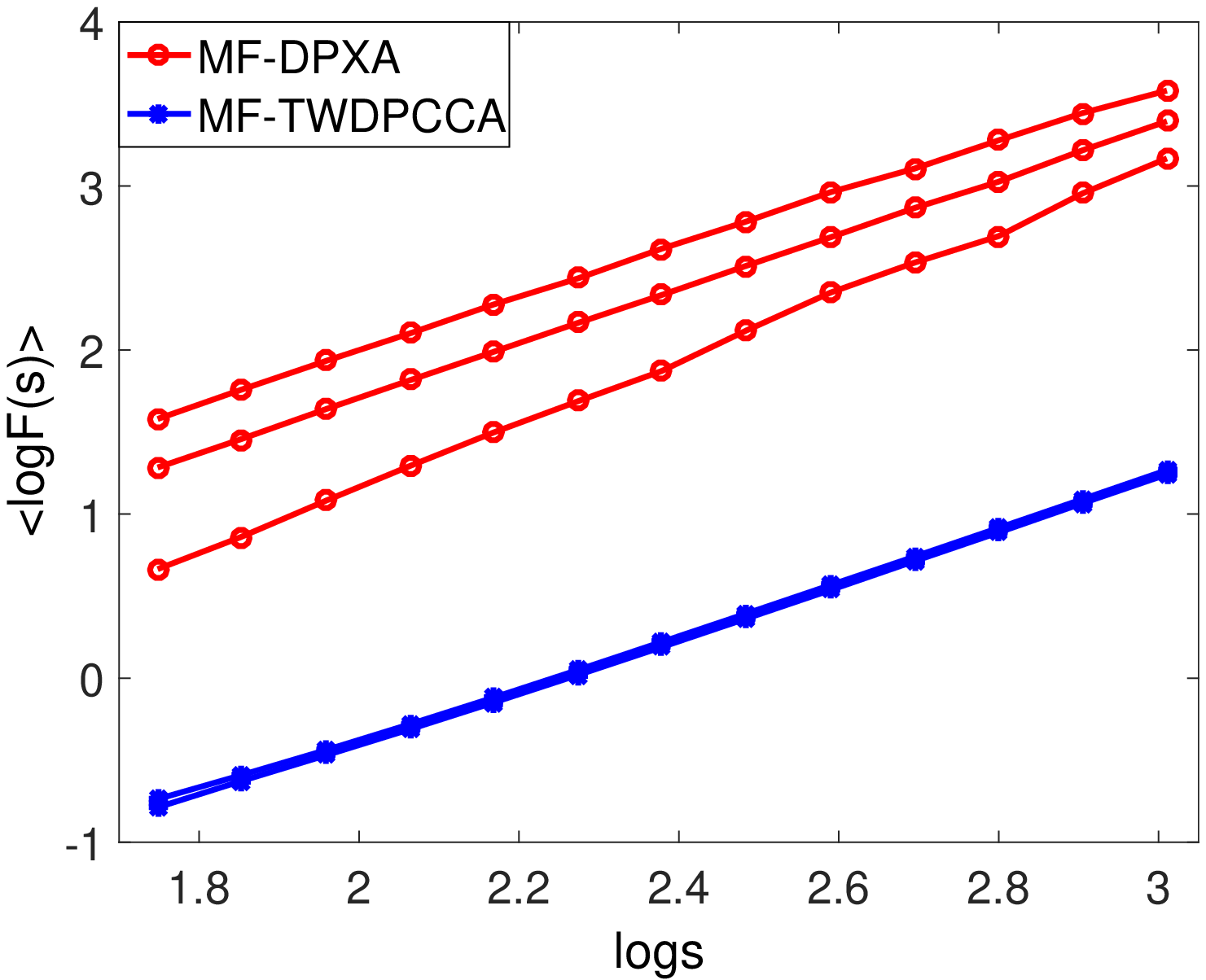}
  \end{minipage}
  \begin{minipage}{7cm}
  \centering
 \includegraphics[width=7cm]{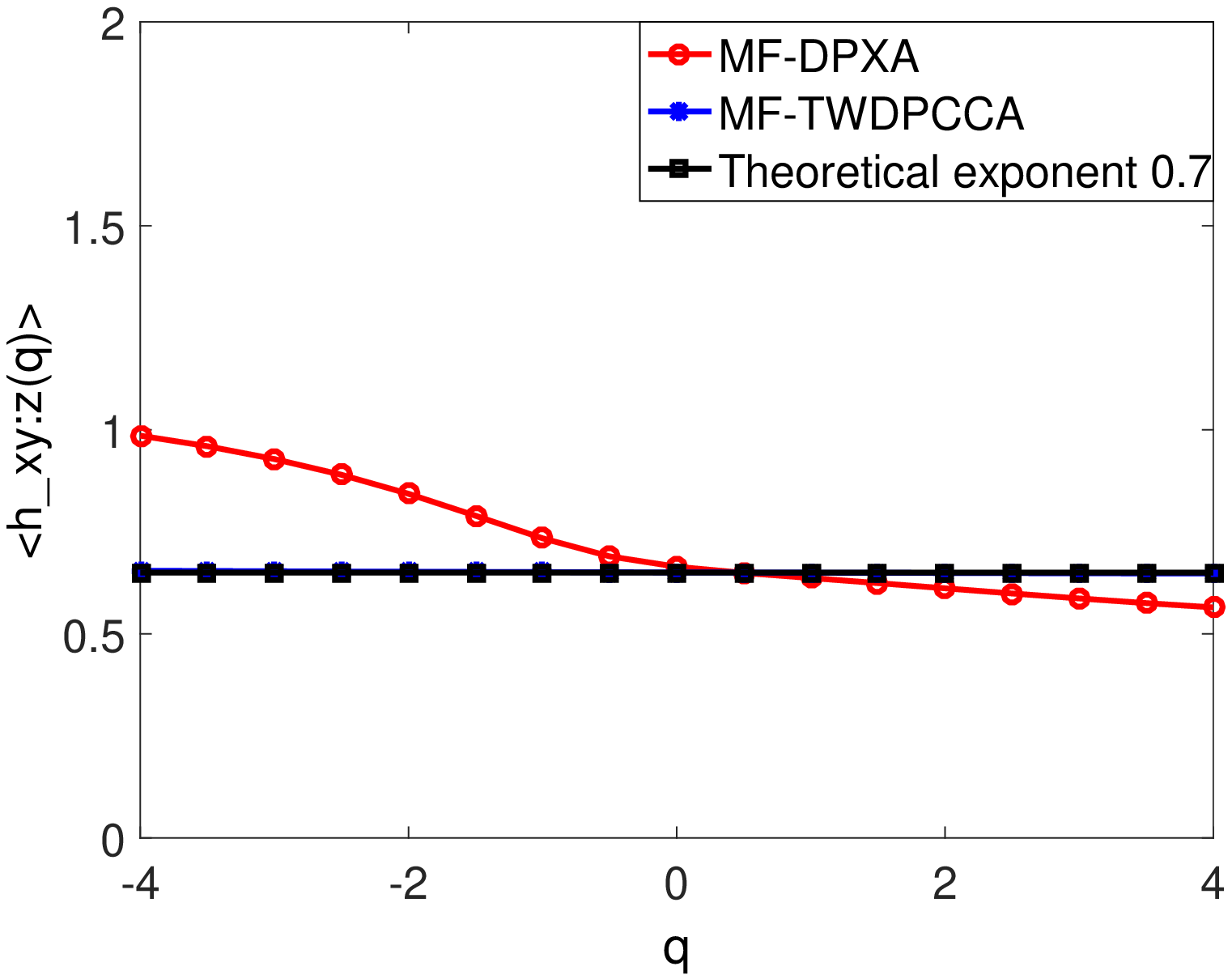}
  \end{minipage}
\text{(a) $H_{r_{x}}=0.6, H_{r_{y}}=0.7,H_{z}=0.5$.}
\\
\centering
  \begin{minipage}{7cm}
  \centering
\includegraphics[width=7cm]{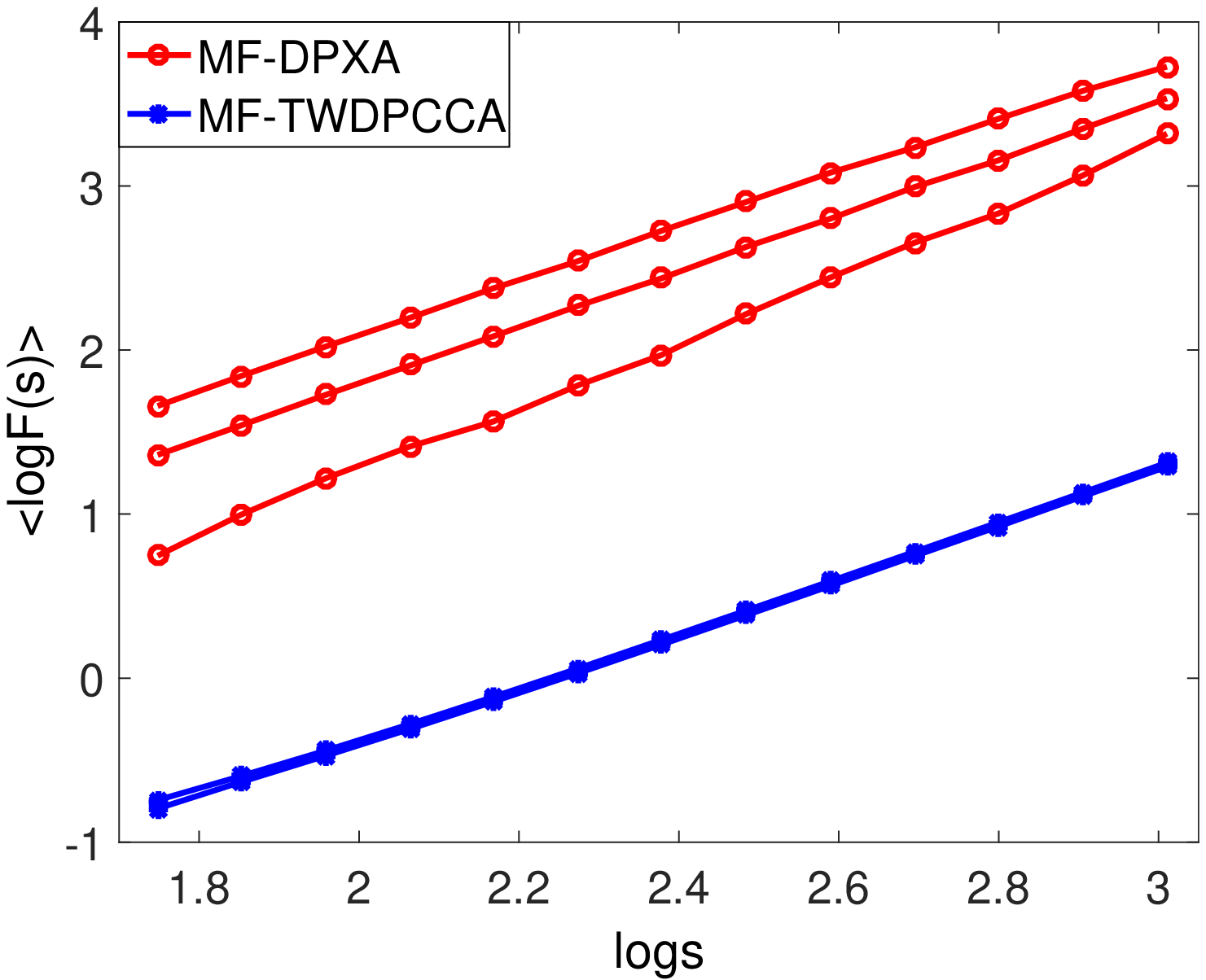}
  \end{minipage}
  \begin{minipage}{7cm}
  \centering
 \includegraphics[width=7cm]{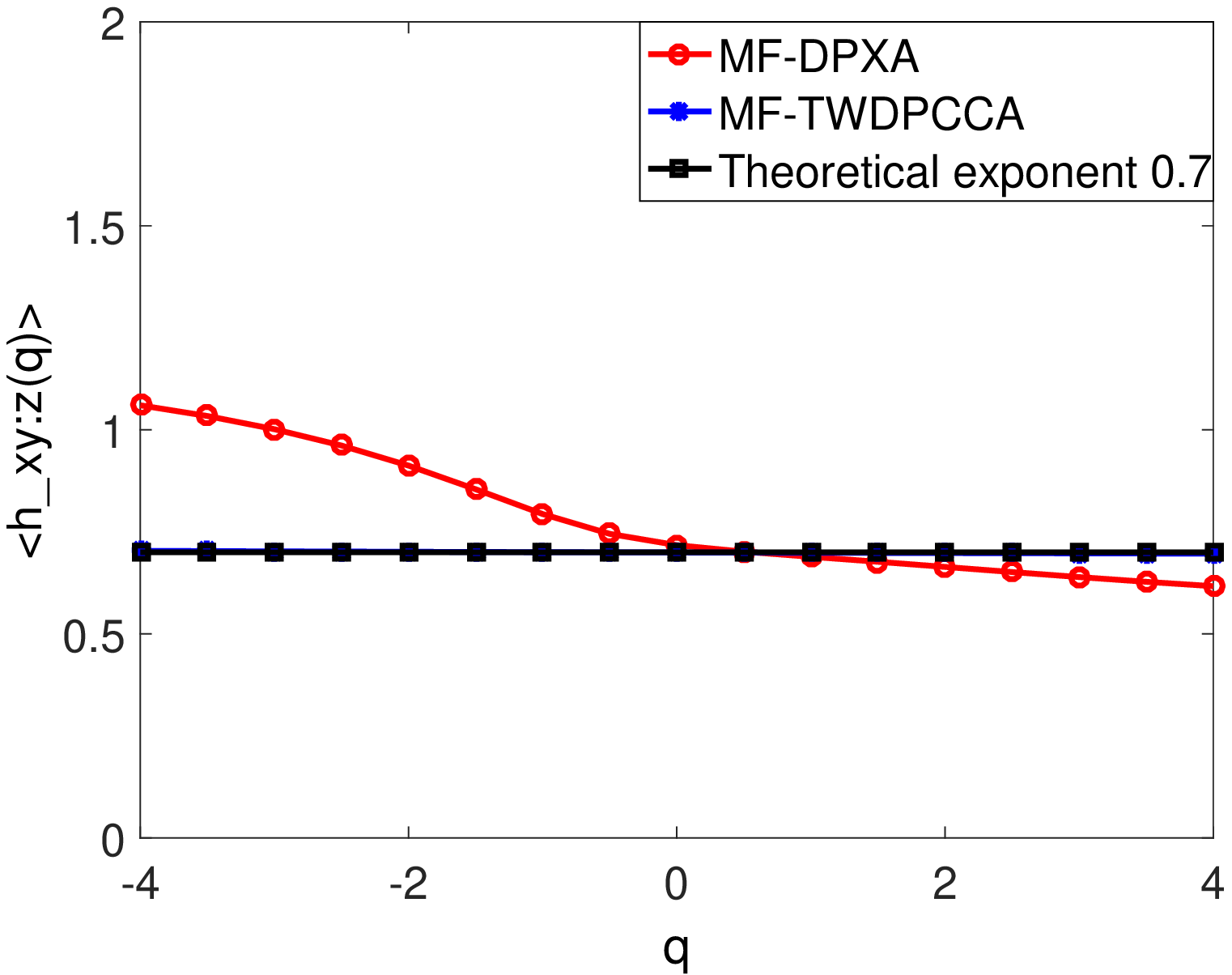}
  \end{minipage}
\text{(b)$H_{r_{x}}=0.6, H_{r_{y}}=0.8,H_{z}=0.5$.}
\\
\centering
  \begin{minipage}{7cm}
  \centering
\includegraphics[width=7cm]{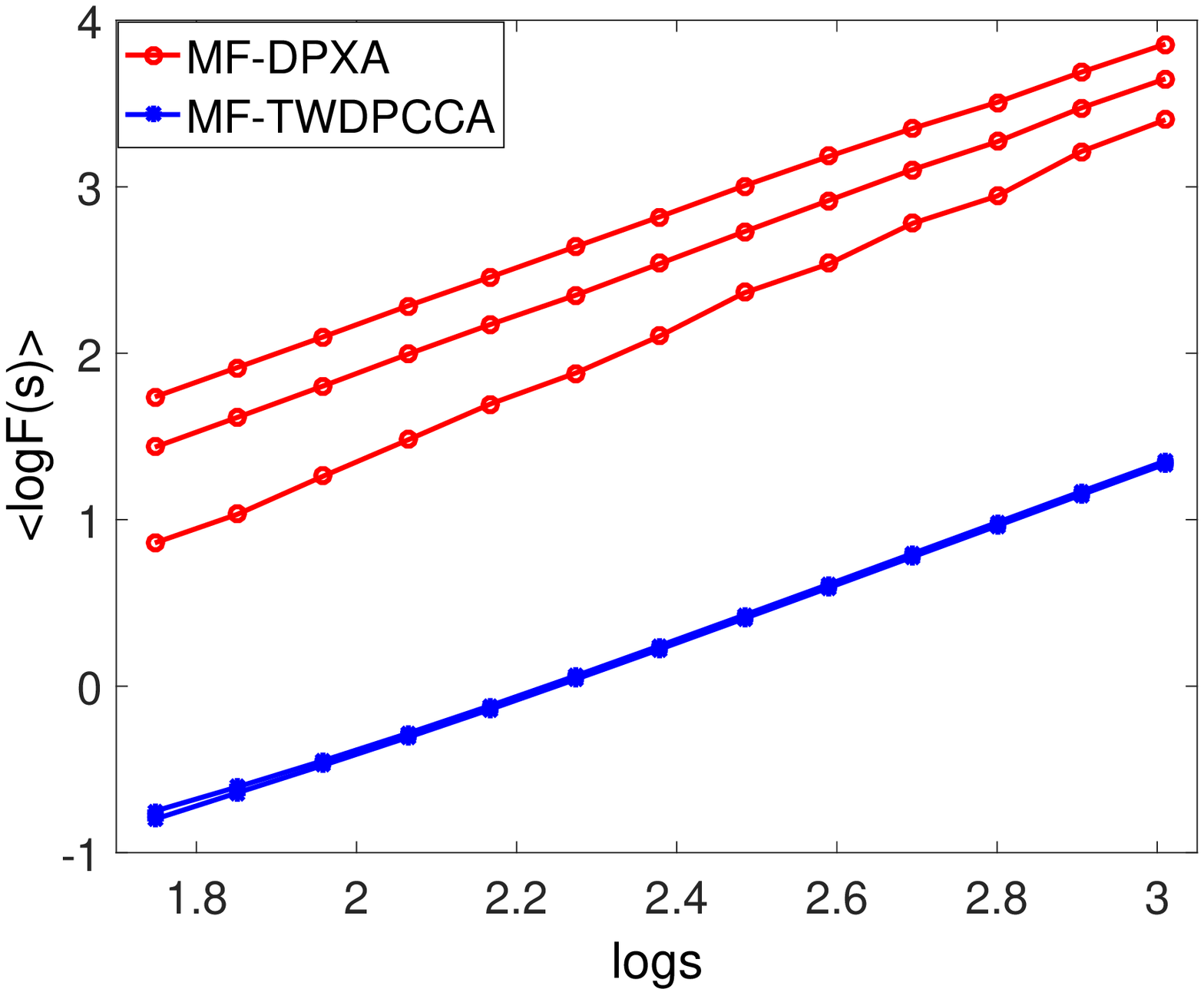}
  \end{minipage}
  \begin{minipage}{7cm}
  \centering
 \includegraphics[width=7cm]{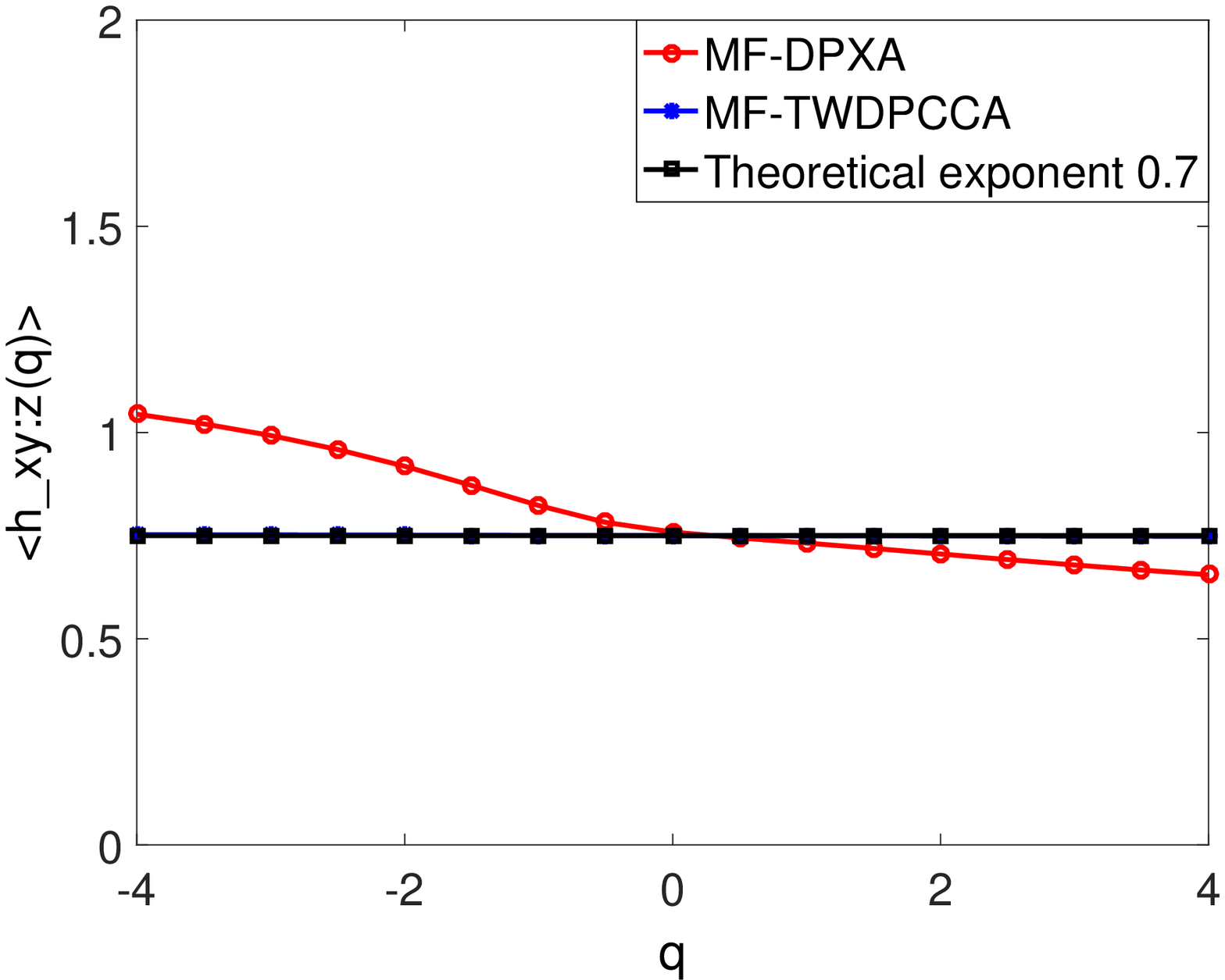}
  \end{minipage}
\text{(c) $H_{r_{x}}=0.6, H_{r_{y}}=0.9,H_{z}=0.5$.}
\caption{The left figures are the power-law dependence of the fluctuation function on the time scale $s$ at $q =-4,0,4$, and the right figures are the scale index $h_{xy:z}(q)$. And  when calculated using MF-TWDPCCA, $c=20$. Here $<\cdot>$ means the average over 100 realizations.}
\label{bfbm}
\end{figure*}

In this section, based on MF-TWXDFA and MF-DPXA, we propose a new method-multifractal temporally weighted detrended partial cross-correlation analysis (MF-TWDPCCA), which can be used to quantify the intrinsic multifractal cross-correlation properties between two non-stationary time series subject to a common external force.

For a given two non-stationary time series recorded simultaneously, $\{x(t):t=1,2,...,N\}$ and $\{y(t):t=1,2,...,N\}$,
 who are affected by time series $\{z(t):t=1,2,...,N\}$, the main steps of MF-TWDPCCA are as follows:

Step 1: First we remove the effect of $z(t)$. The additive model for  models of $x(t)$ and $y(t)$ can be given respectively as:
\begin{eqnarray}
\left\{\begin{array}{ll}
x(t)=\beta_{x,0}+\beta_{x,1}z(t)+r_{x}(t),\\
y(t)=\beta_{y,0}+\beta_{y,1}z(t)+r_{y}(t).
\end{array}\right.\label{linearr}
\end{eqnarray}
where, $t=1,2,...,N$. When using regression analysis to find the residual sequence, in a window of length $s$, we use the idea in MF-TWXDFA \cite{WeiYL2017} to remove the effect of the sequence $z(t)$ on $x(t)$ and $y(t)$ point by point as follows. For a given integer $c$ ($c\ge 2$), the points $j$ contained in a sliding window $MW_i$ corresponding to the point $i$ should satisfy $|i-j|\leq [\frac{s}{c}]$. When the length of time series is different, we take different value for $c$. Usually the value of $c$ is determined by experience. Accordingly, the weight function of the geographic weighted regression model is:
\begin{eqnarray}
\omega_{ij}=
\left\{\begin{array}{ll}
[1-(\frac{c(i-j)}{s})^{2}]^{2}, & {\mbox if}\ |i-j|\leq[\frac{s}{c}],\\
 0, &  {\mbox otherwise.}
 \end{array}\right.\label{weighte}
\end{eqnarray}

In the window $MW_i$, We perform linear regression for $\{\omega_{ij}x_j\}$ on $\{z_j\}$ or $\{\omega_{ij}y_j\}$ on $\{z_j\}$, respectively. We can get the regression values $\hat{x}(z_{i})$ and $\hat{y}(z_{i})$ of $x(i)$ and $y(i)$, respectively. Then we get the corresponding residual sequence:
\begin{eqnarray}\nonumber
\begin{array}{ll}
\hat{r}_{x}(i)=x(i)-\hat{x}(z_{i})\\
\hat{r}_{y}(i)=y(i)-\hat{y}(z_{i}).
 \end{array}
\end{eqnarray}

 Step 2: For the newly obtained residual sequences $\{\hat{r}_x(t):t=1,2,...,N\}$ and $\{\hat{r}_y(t):t=1,2,...,N\}$, calculate their cumulative dispersion respectively:
\begin{eqnarray}\nonumber
\begin{array}{ll}
R_{x}(i) = \sum\limits_{t = 1}^i(\hat{r}_{x}(t)-\overline{r_{x}})\\
 R_{y}(i) = \sum\limits_{t = 1}^i(\hat{r}_{y}(t)-\overline{r_{y}}),
 \end{array}
\end{eqnarray}
where $i=1,2,\cdots,N$ and $\overline{r_{x}}=\frac{1}{N}\sum\limits_{t = 1}^N \hat{r}_{x}(t),\overline{r_{y}}=\frac{1}{N}\sum\limits_{t = 1}^N\hat{r}_{y}(t)$.

 Step 3: Split $R_{x}(i),R_{y}(i)$ into ${N_s}$ non-overlapping intervals of length $s$, where
  $ N_s = [N/\mathord s]$. Considering that the sequence length may not be an integer multiple of $ s $, in order to make full use of all data and avoid losing the tail data information, the $R_{x}, R_{y}$ are divided twice, that is, once from front to back and from back to front, and then we get ${2N_s}$ subintervals.

 Step 4: Calculate the local trends $\hat{R}_{x}(i)$ and $\hat{R}_{y}(i)$ of $R_{x}(i)$ and $R_{y}(i)$ in window $MW_i$ by the geographic weighted regression method as MF-TWXDFA \cite{WeiYL2017}.

 Step 5: For the $v$th interval, calculate the detrended partial  cross-correlation fluctuation function $f_{xy:z}(v,s):$
\begin{align}
\centering
f_{xy:z}(v,s)=& \frac{1}{s}\sum\limits_{i=1}^s\{[R_{x}(m_{i})-\hat{R}_{x}(m_{i})][R_{y}(m_{i})-\hat{R}_{y}(m_{i})]\},~~~~\nonumber\\
   \text{where } m_{i}=&\left( {v - 1} \right)s + i, v=1,2,\cdots,N_{s};\nonumber\\
f_{xy:z}(v,s)=& \frac{1}{s}\sum\limits_{i=1}^s\{[R_{x}(m^{*}_{i})-\hat{R}_{x}(m^{*}_{i})][R_{y}( m^{*}_{i}) - \hat{R}_{y}(m^{*}_{i})]\},~~~\nonumber\\
\text{where } m^{*}_{i}=&N - \left( {v - {N_s}} \right)s + i,v = {N_s} + 1,\cdots ,2{N_s}.\nonumber
\end{align}

Step 6: For the ${2N_s}$ intervals, calculate the average of $f_{xy:z}(v,s)$ to get the $q$ order fluctuation function
$F_{xy:z}(q, s)$:
\begin{eqnarray}
&F_{xy:z}(q,s)=
&\Big{|}\frac{1}{{{2N_s}}}\sum\limits_{v = 1}^{{2N_s}} {sgn({f}_{xy:z}\left( {v,s} \right))\left| {{f}_{xy:z}\left( {v,s} \right)} \right|}  ^{{q \mathord{\left/{\vphantom {1 2}} \right. \kern-\nulldelimiterspace} 2}}\Big{|}^{\frac{1}{q}} ,\nonumber
\text{for}\ q\neq0.
\end{eqnarray}
\begin{eqnarray}
&F_{xy:z}(q,s)=
&exp\Big{[}\frac{1}{4N_{s}}\sum\limits_{v = 1}^{{2N_s}}sgn({f}_{xy:z}\left( {v,s} \right))ln|{f}_{xy:z}\left( {v,s} \right)|\Big{]} ,\nonumber
\text{for}\ q=0.
\end{eqnarray}

Step 7: Determine the scale index of $F_{xy:z}(q,s)$. If there is a long-range cross-correlation between the time series $x(t)$ and $y(t) $, then $F_{xy:z}(q,s)$ satisfies the following power law relationship:
\[F_{xy:z}\left( q,s \right) \sim {s^{h_{xy:z}(q)}}.\]

As we know, the scale index $h_{xy:z}(q)$ is a fractal exponent that quantifies the intrinsic cross-correlation between $x(t)$ and $y(t)$. If $h_{xy:z}(q)$ is independent of $q$, the intrinsic cross-correlation of $x(t)$ and $y(t)$ corresponds to single fractal case; If $h_{xy:z}(q)$ changes with the change of $q$, the intrinsic cross-correlation of $x(t)$ and $y(t)$ corresponds to multifractal case\cite{WangF2016}.

According to the standard multifractal formalism, multifractal properties can also be characterized by the multifractal mass exponent $\tau(q)$, i.e.,
\[\tau_{xy:z}(q)=qh_{xy:z}(q)-D_f,\]
where $D_f$ is the fractal dimension of the geometric support of the multifractal measure\cite{Kantelhardt2002}.

%%%%%%%%fig3
\begin{figure*}[!htb]
\centering
  \begin{minipage}{8cm}
  \centering
  \includegraphics[width=8cm]{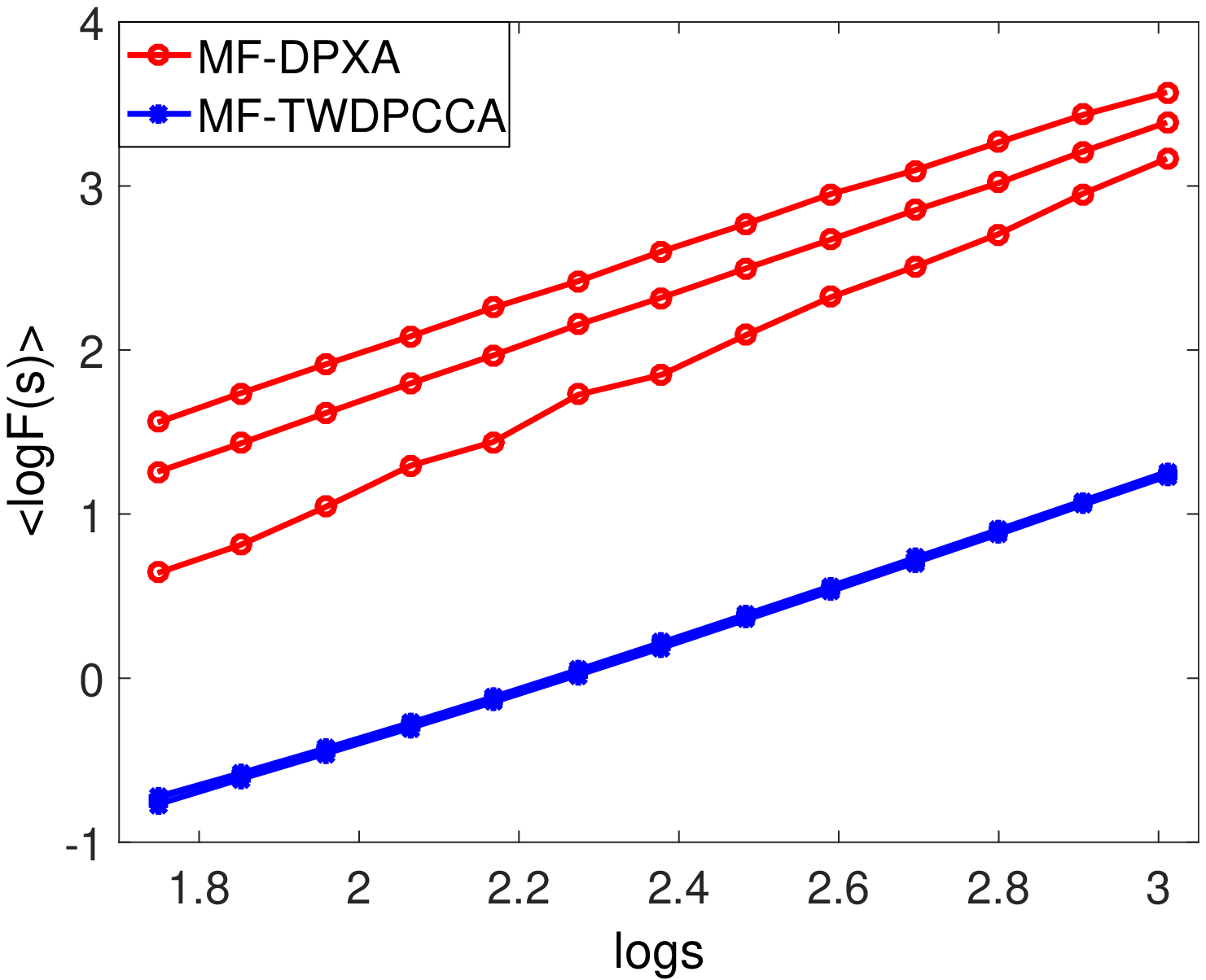}
  \end{minipage}
  \begin{minipage}{8cm}
  \centering
  \includegraphics[width=8cm]{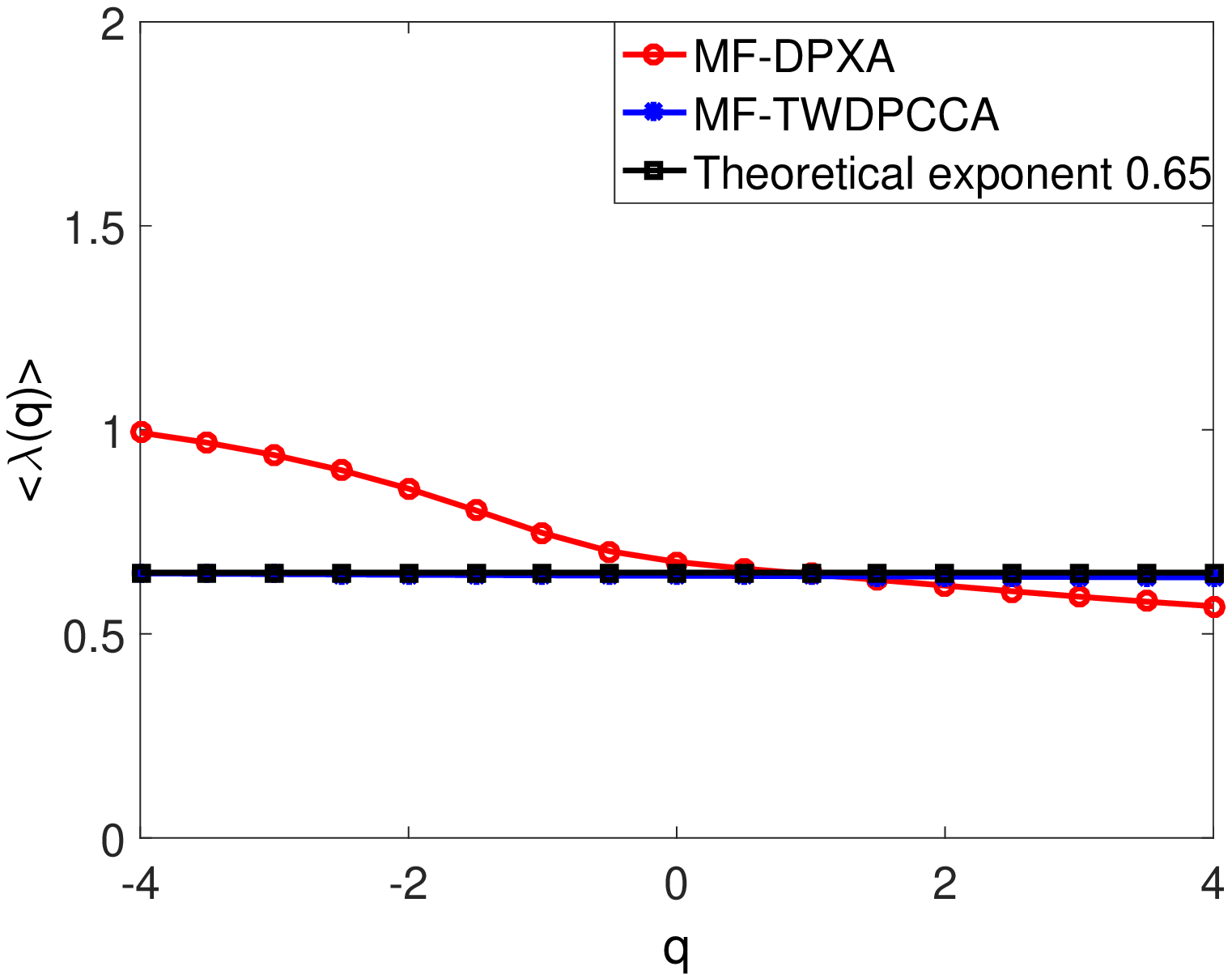}
  \end{minipage}
\caption{ $H_{r_{x}}=0.6 $, $ H_{r_ {y}}=0.7$, $H_{z}=0.8$. Similarly, the left figure is the power-law dependence of the fluctuation function on the time scale $s$ at $q =-4,0,4$, and the right figure is the scale index $h_{xy:z}(q)$. And  when calculated using MF-TWDPCCA, $c=20$.  Here $<\cdot>$ means the average over 100 realizations.}\label{bfbmadd}
\end{figure*}

%%%%%%%%fig4
\begin{figure*}[!htb]
\centering
  \begin{minipage}{8cm}
  \centering
  \includegraphics[width=8cm]{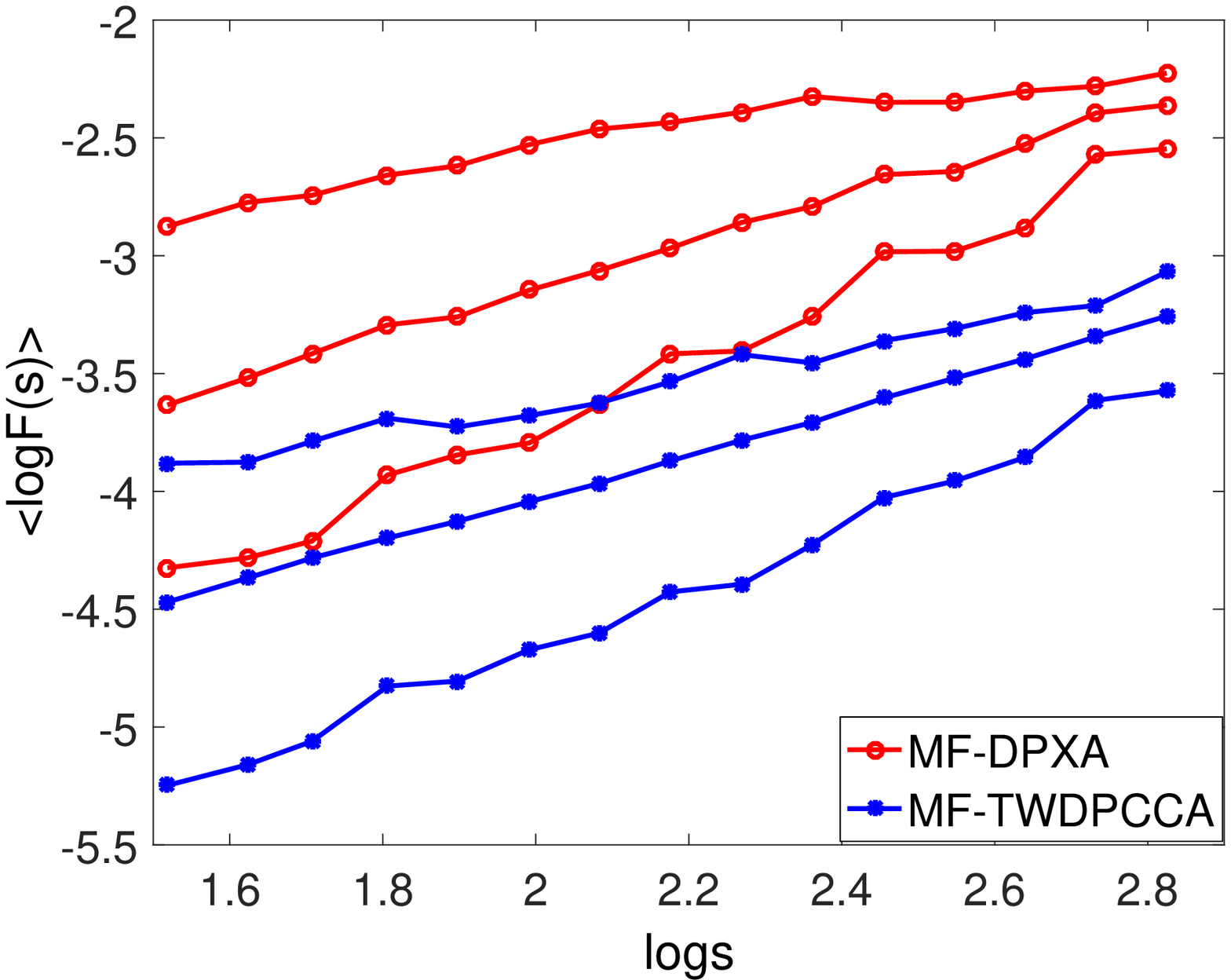}
  \end{minipage}
  \begin{minipage}{8cm}
  \centering
  \includegraphics[width=8cm]{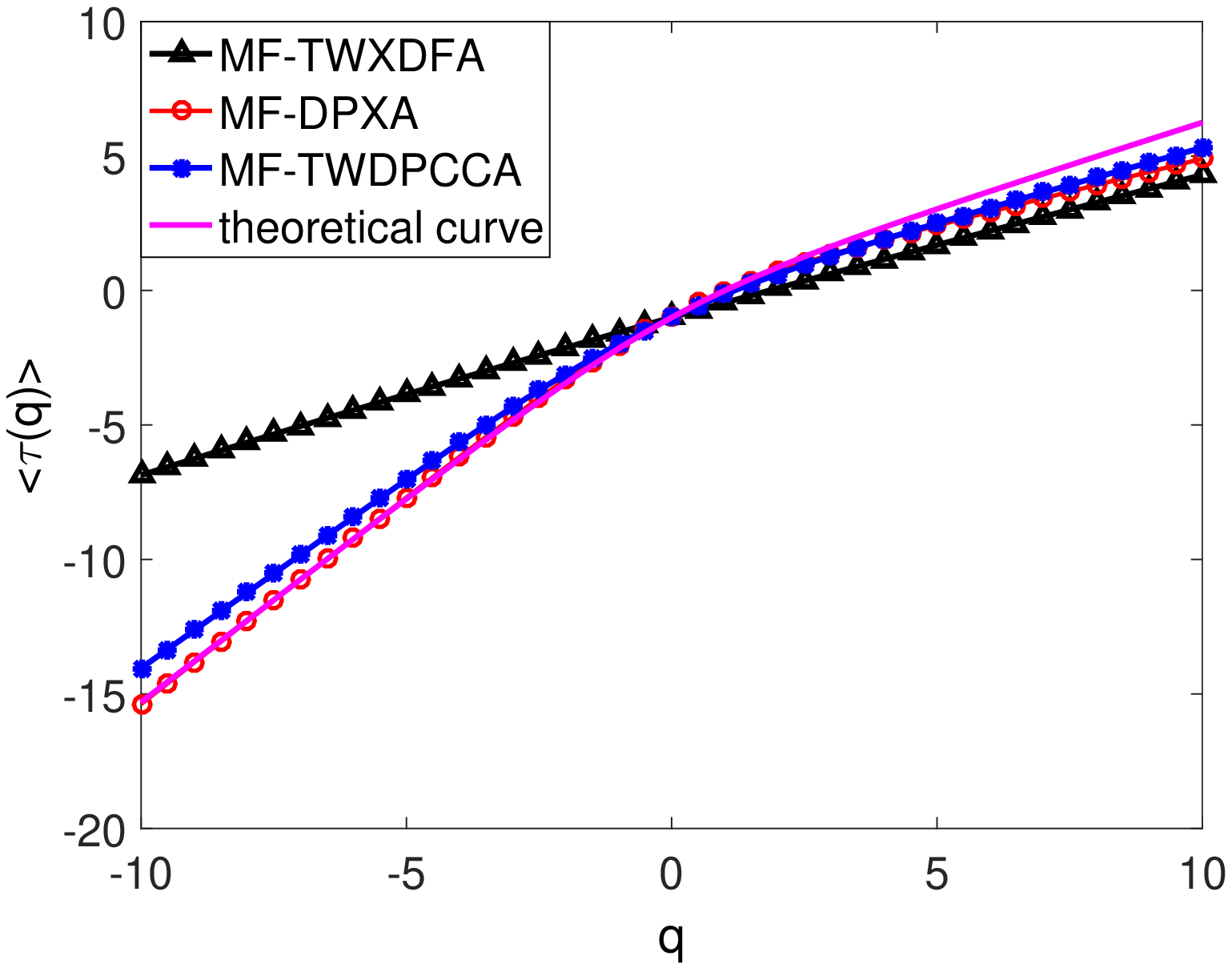}
  \end{minipage}
\caption{The left figure shows the power-law dependence of the fluctuation function $F_{xy:z}(q,s)$ on the scale $s$ for $q=-4,0$, and $4$. The right figure shows the $\tau_{xy}(q)$, $\tau_{xy:z}(q)$ by MF-TWXDFA, MF-TWDPCCA, MF-DPXA, and the theoretical curve $\tau_{r_xr_y}$ is shown as a continuous line. Here $<\cdot>$ means the average over 100 realizations. }\label{bMfspictures}
\end{figure*}

%%%%%%%%%%%%%% fig:5插值：互相关和偏互相 logF(s) 对比
\begin{figure}[!htb]
\centering
\includegraphics[width=10cm,height=8cm]{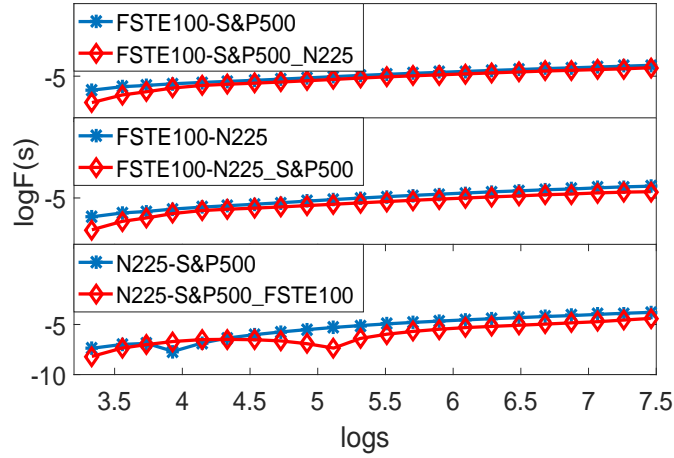}
\vspace{-2em}\caption{The fluctuation function $logF(s)$ calculated by MF-TWXDFA and MF-TWDPCCA. }\label{vs_of_logF(s)}
\end{figure}

%%%%%%%%%%%%%%fig:6 插值：互相关和偏互相相关 rho(s) 对比
\begin{figure}[!htb]
\centering
\includegraphics[width=10cm,height=8cm]{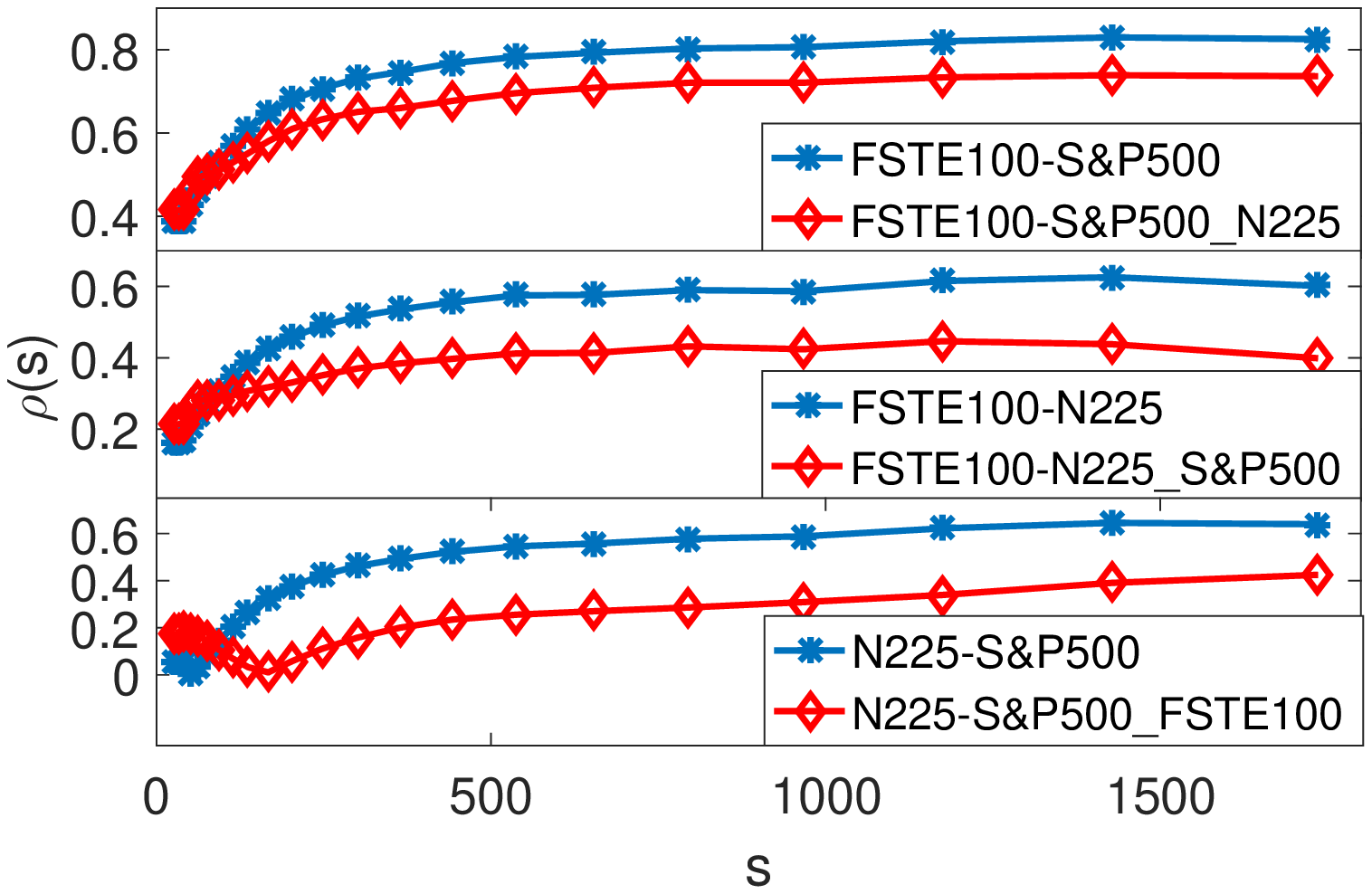}
\vspace{-2.5em}\caption{The $\rho(s)$ calculated by MF-TWXDFA and MF-TWDPCCA.}\label{vs_of_rho(s)}
\end{figure}

%%%%fig:7插值：互相关和偏互相相关 h(q) 对比

\begin{figure}[!htb]
\centering
\includegraphics[width=10cm,height=8cm]{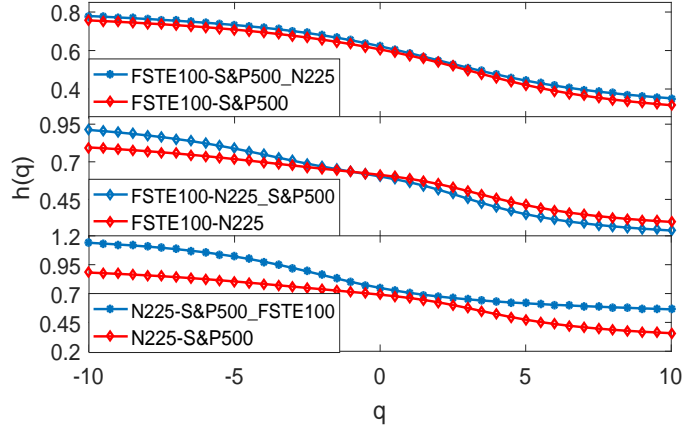}
\vspace{-3em}\caption{ The $h(q)$ calculated by MF-TWXDFA and MF-TWDPCCA}\label{vs_of_h(q)}
\end{figure}

\section{Numerical experiments}

\ \ In order to evaluate the performance of MF-TWDPCCA, in this section, we use the additive models of $x$ and $y$ as Eq. (1) to perform numerical simulation and verify the effectiveness of our method.
%\begin{eqnarray}
%\begin{array}{ll}
%x(t) = \beta _{x,0} + \beta _{x,1}z(t) + r_x(t)\\
%y(t) = \beta _{y,0} + \beta _{y,1}z(t) + r_y(t).
% \end{array}\label{addmodel}
%\end{eqnarray}

In the experiments, we repeated the experiment $100$ times. The sequence length was selected as $ 2 ^ {12} $. After calculation and comparison, we take $c = 20$ in this work.

\subsection{TWDPCCA coefficient}

 \ \ We define a new partial correlation coefficient (TWDPCCA coefficient) which is very similar to the DPXA coefficient in \cite{QianXY2015}. Firstly we define:

\begin{eqnarray} \nonumber
\begin{array}{ll}
F_{xy:z}(2,s)=\Big{|}\frac{1}{{{2N_s}}}\sum\limits_{v = 1}^{{2N_s}} {sgn({f}_{xy:z}\left( {v,s} \right))\left| {{f}_{xy:z}\left( {v,s} \right)} \right|}\Big{|}^{1/2} \\
F_{x:z}(2,s)=\Big{|}\frac{1}{{{2N_s}}}\sum\limits_{v = 1}^{{2N_s}} {sgn({f}_{xx:z}\left( {v,s} \right))\left| {{f}_{xx:z}\left( {v,s} \right)} \right|}\Big{|}^{1/2}\\
 F_{y:z}(2,s)=\Big{|}\frac{1}{{{2N_s}}}\sum\limits_{v = 1}^{{2N_s}} {sgn({f}_{yy:z}\left( {v,s} \right))\left| {{f}_{yy:z}\left( {v,s} \right)} \right|}\Big{|}^{1/2}.
 \end{array}
\end{eqnarray}
Then we get
$$\rho_{TWDPCCA}(s)=\rho_{xy:z}(s)=\frac{{F_{xy:z}^2{(2,s)}}}{F_{x:z}(2,s) \cdot F_{y:z}(2,s)}$$
It is easy to find that $\rho_{TWDPCCA}(s)$ satisfies $-1\leq\rho_{TWDPCCA}(s)\leq 1$.

We use a mathematical model in equation (1) for numerical simulation, where $z(t)$ is a fractal Gaussian noise with $Hurst$  index $H_{z}$, and $r_{x}$ and $r_{y}$ are the incremental series of the two components of a bivariate fractional Brownian motion ({\bf BFBMs}) with Hurst
indices $H_{r_x}$ and $H_{r_y}$\cite{VenugopalV2006,YuZG2007,YuZG2009}. In the simulations we set $H_{r_{x}}=0.1, H_{r_{y}}=0.1,\rho=0.7, H_{z}=0.95, \beta _{x,0}=\beta _{y,0}=2,\beta _{x,1}=\beta _{y,1}=3$, where $\rho$ is the  cross-correlation
coefficient between $r_{x}$ and $r_{y}$. And then we generate $100$ sets of sequences of length $2^{12}$ to verify the validity of {\bf TWDPCCA}.

As shown in Figure \ref{rhocompaerd}, because $z$ has a strong effect on $r_{x}$ and $r_{y}$, the performance of $x$ and $y$ is dominated by $z$. The cross-correlation coefficients $\rho_{xy}$ calculated by both MF-TWXDFA and MF-DCCA are all close to $1$ especially when the window length $s$ is relatively large. Compare left part and right part of Figure 1, we could find that the partial correlation coefficients $\rho_{r_xr_y}$ calculated by MF-TWDPCCA are more accurate than that calculated by MF-DPXA. In addition, MF-TWDPCCA can be applied to a wider window length.

\subsection{ Bivariate Fractional Brownian Motion (BFBMs)}

\ \ In this section, in order to further test the performance of MF-TWDPCCA, we use it to calculate the multifractal properties of BFBMs in the three sets of the above additive models (Eq. (1)). Same as in the previous section, $z(t)$ is a fractal Gaussian noise with $Hurst$  index $H_{z}$, and $r_{x}$ and $r_{y}$ are the incremental series of the two components of BFBMs with Hurst indices $H_{r_x}$ and $H_{r_y}$. Extensive research on BFMS has been made. We know that BFBMs is a single fractal process and there is a relationship $H_{{r_{x}}{r_{y}}}=(H_{r_{x}}+ H_ {r_{y}})/2$ \cite{LavancierF2009,CoeurjollyJF2010,AmblardPO2011}. Here $H_{{r_{x}}{r_{y}}}$ is used to represent the theoretical value of the cross-correlation coefficient between $r_{x}$ and $r_{y}$. In the simulations, the length of all the sequences that we generated is $2^{12}$ and we set: $\beta _{x,0}=\beta _{y,0}=2,\beta _{x,1}=\beta _{y,1}=3,H_z=0.5$; $(a)H_{r_{x}}=0.6, H_{r_{y}}=0.7;\  (b)H_{r_{x}}=0.6, H_{r_{y}}=0.8;\  (c)H_{r_{x}}=0.6, H_{r_{y}}=0.9$.

From the left figures in Figure \ref{bfbm}, we know that the relationships calculated by both MF-TWDPCCA and MF-TWDPXA  between the fluctuation functions and the time scale $s$ are all approximately power law relationship, which shows that BMBFs is fractal process. And within a certain error range, each fitted straight line is approximately parallel, which further indicates that BMBFs is a single fractal process. However, we can also find that compared with MF-DPXA, MF-TWDPCCA can get a smoother logarithmic plot, which is more accurate in fitting the results.

We denote the maximum difference $max(h_{xy: z}(q))-min(h_{xy:z}(q))$ calculated by MF-TWDPCCA as$\triangle h_{xy:z}(q)$.
For the above 3 cases, the maximum and minimum values of $h_{xy:z}$ are $0.655\&0.648,0.704 \\
\&0.697,0.752\&
0.748$, respectively. So $\triangle h_{xy:z}$ are $0.007, 0.007$ and $0.004$. These values are relatively small, indicating that the fluctuation of $h_{xy:z}$  is small and it is approximately independent on $q$. On the other hand, from the right figures in Figure \ref{bfbm}, we can find that the the scale index $h_{xy:z}(q)$ calculated by MF-TWDPCCA are all approximate straight lines. All these shows that BFBMs is a single fractal process, which is consistent with the fact. However, there is some fluctuation in $h_{xy:z}(q)$ obtained by MF-DPXA.

In order to further verify the applicability of MF-TWDPCCA, we select the $Hurst$ index as $H_{r_{x}}=0.6 $, $ H_{r_ {y}}=0.7$, $H_{z}=0.8$ for a set of sequences to perform the same numerical simulation. From Figure \ref{bfbmadd}, where the $ Hurst $ index of the fractal Gaussian noise is $0.8$, we can find that MF-TWDPCCA still get better results.

\subsection{Multifractal binomial measures}

\ \ As we all know, the binomial measures\cite{YuZG2014,BaoJiangZ2007} produced by the $p$ model have multifractal properties and its sacale index $\tau(q)$function is known. We combine them with Gaussian noise to test the performance of MF-TWDPCCA. Two binomial measures $\{r_x(t):t = 1,2,...,2^{12}\}$ and $\{r_y(t):t = 1,2,...,2^{12}\}$ are generated iteratively\cite{WeiYL2017} by using corresponding probability parameters $p_x=0.3$ for $r_x$ and $p_y=0.4$ for $r_y$ respectively. In the additive model (Eq.(1)) we set $\beta _{x,0}=\beta _{y,0}=2,\beta _{x,1}=\beta _{y,1}=3$, and then we get contaminated signals $x=2+3z+r_x$ and $y=2+3z+r_y$, where $z$ is Gaussian noise with Hurst index $H_z=0.5$.

As shown in Figure \ref{bMfspictures}, for the binomial measure, there is no obvious difference between the logarithmic plots obtained by MF-TWDPCCA and MF-DPXA. And these two methods get almost the same $\tau(q)$, especially when $q>0$. All these show that the two methods can get almost the same results for binomial measures. However, when using MF-TWXDFA to analyze the cross-correlation between $x$ and $y$, without removing the influence of the common potential factor $z(t)$, the multifractal quality index function is approximately a straight line, which fails to uncover any multifractality between $r_{x}(t), r_{y}(t)$. This shows that the multifractal partial cross-correlation analysis is necessary.

\section{Application to stock market index}

 \ \ In this section, we apply the MF-TWDPCCA proposed by us for empirical research. Daily closing data for three stock market indexes, which are FSTE100, S$\&$P500 and N225 (finace.yahoo.com database) from January 4, 2001 to January 9, 2019, are used. For the dates without data record, we use the data of five days before and after them to do linear interpolation to complete the data of that day. Here we also take $c=20$. The daily rate of return is calculated as follows : $R_t=ln(P_t)-ln(P_{t-1})$, where $P_t$ is the closing price on the $t$th days. We can easily find that the mean of all three yield data is close to 0 and less than their standard deviation, which indicates that both low returns and high risks coexist.

 Figures \ref{vs_of_logF(s)}, \ref{vs_of_rho(s)} and  \ref{vs_of_h(q)} show the results calculated by us, where
 "FSTE100-S$\&$P500, FSTE100-N225 and N225-S$\&$P500" means they are calculated by MF-TWXDFA, and "FSTE100-S$\&$P500\_N225, FSTE100-N225\_S$\&$P500 and N225-S$\&$P500\_FSTE100" means they are calculated by MF-TWDPCCA. From Figure \ref{vs_of_rho(s)}, we can find that the cross-correlation coefficients between the three sequences studied are overestimated in most cases of scale $s$. Figure \ref{vs_of_h(q)} shows that there exists multifractal cross-correlation among the three sequences studied.

\section{Discussion and conclusions}

\ \ In this paper, based on the idea of MF-TWXDFA and MF-DPXA, for studying the intrinsic cross-correlation between two non-stationary time series affected by common external factors, we propose a novel method---Multifractal temporally weighted detrended partial cross-correlation analysis (MF-TWDPCCA). In the numerical experiments, we compared MF-TWDPCCA with MF-DPXA. Firstly, as shown in Figure \ref{rhocompaerd}, we find that when calculating the detrended partial cross-correlation coefficients of the additive model (Eq. (1)), both MF-TWDPCCA and  MF-DPXA reveal the intrinsic relationship between the two time series. However MF-TWDPCCA obtains more accurate results while having a wider range of sliding window length $s$. Secondly, as shown in Figure \ref{bfbm} and Figure \ref{bfbmadd}, when calculating the bivariate fractional Brownian motion with unifractal properties, MF-TWDPCCA not only obtains a smoother bi-logarithmic power law graph, but also obtains the single fractal property which are almost consistent with the theoretical results. And then, comparing the performance of two methods in computing multifractal binomial measures affected by white noise, we find both MF-TWDPCCA and MF-DPXA obtain the intrinsic multifractal relationship over a certain length $s$ between the two time series, and the results between the two methods are not significantly different. Finally, we applied MF-TWDPCCA to real stock market data, as show in Figures \ref{vs_of_rho(s)} and \ref{vs_of_h(q)}, we find that there exists multifractal cross-correlation among the three sequences studied. The cross-correlation coefficients between the three sequences studied by MF-TWXDFA are overestimated in most cases of scale $s$. So when we study the correlation between two sequences, if we find that they receive the influence of common external force from the mechanism of their internal background, it is necessary to conduct partial cross-correlation analysis. In a word although MF-TWDPCCA is more time-consuming than MF-DPXA, MF-TWDPCCA can better reveal the intrinsic relationship between two non-stationary time series. At last, we think that we can get more reliable results when we use MF-TWDPCCA to study intrinsic cross correlation between non-stationary time series affected by common external factors.

\section*{Acknowledgement}

This project was supported by the National Natural Science Foundation of China (Grant No. 11871061),
Collaborative Research project for Overseas Scholars (including Hong Kong and Macau) of the National Natural Science Foundation of China (Grant No. 61828203), the Chinese Program for Changjiang Scholars and Innovative Research Team in University (PCSIRT) (Grant No. IRT$\_$15R58) and the Hunan Provincial Innovation Foundation for Postgraduate (Grant No. CX2017B265).

\nocite{*}
%\bibliography{aipsamp}% Produces the bibliography via BibTeX.

\end{document}